\documentclass[10pt,conference]{IEEEtran}
\usepackage{cite}
\ifCLASSINFOpdf
   \usepackage[pdftex]{graphicx}
\else
\fi
\usepackage{amsmath}
\usepackage{algorithmic}
\usepackage{array,multirow}

\begin{document}
\title{Optimization in Large Graphs: Toward a Better Future?}

\author{\IEEEauthorblockN{Pieter Leyman \& Patrick De Causmaecker}
\IEEEauthorblockA{CODeS research group, Department of Computer Science \& imec-ITEC \\KU Leuven - KULAK\\ 
Etienne Sabbelaan 53, 8500 Kortrijk (Belgium)\\
Email: pieter.leyman@kuleuven.be, patrick.decausmaecker@kuleuven.be}}

\maketitle

\begin{abstract}
Finding groups of connected individuals in large graphs with tens of thousands or more nodes has received considerable attention in academic research. In this paper, we analyze three main issues with respect to the recent influx of papers on community detection in (large) graphs, highlight the specific problems with the current research avenues, and propose a first step towards a better approach. 

First, in spite of the strong interest in community detection, a strong conceptual and theoretical foundation of connectedness in large graphs is missing. Yet, it is crucial to be able to determine the specific feats that we aim to analyze in large networks, to avoid a purely black-or-white view. 

Second, in literature commonly employed (meta)heuristic frameworks are applied for the large graph problems. Currently, it is, however, unclear whether these techniques are even viable options, and what the added value of the constituting parts is. Additionally, the manner in which different algorithms are compared is also ambiguous.

Finally, no analyses of the impact of data parameters on the reported clusters is done. Nonetheless, it would be interesting to evaluate which characteristics lead to which type of communities and what their effect is on computational difficulty.
\end{abstract}

\IEEEpeerreviewmaketitle

\section{\label{intro}Introduction}
Optimization problems are everywhere in our daily lives. As an example, consider a warehouse of a third party logistics (3PL) service provider. The 3PL has to determine among others which types of products are stored at which location in the warehouse, how orders should be picked and the quantities of products stored. The goal of the company is to optimize one or more objectives such as minimize throughput time or maximize space usage, while taking limitations such as the available work force and storage space into account. 

In practice, the aforementioned problem may involve thousands or even tens of thousands of different product types. In academic research, however, the datasets are often smaller in size. Fortunately, some problems such as large graph problems, with tens of thousands or even million nodes, have been investigated in more detail in recent years. Since graphs underlie many optimization problems (e.g. vehicle routing, project scheduling and water distribution), this is a worthwhile research avenue. Algorithms for these types of problems, however, need to be able to handle the large datasets inherent to such problems. 

Finding groups of individuals connected to a predefined degree in large graphs has received a large amount of attention in literature \cite{fortunato2010}, \cite{fortunato2016}. Examples are the analysis of social media networks such as Facebook and Twitter, and neural pathways in the human brain. All sorts of (meta)heuristic solution approaches have been proposed and analyzed in a great many research papers. We, however, believe that several shortcomings exist in the recent articles on community detection in large graphs. 

\section{\label{issue1}What is a community?}
First, we have to decide what good and bad clusters of nodes are. A proper definition of a community is, however, missing. A complete model with an unambiguous objective function is required to ensure that the problem which we want to solve is perfectly clear. As stated by \cite{fortunato2010}, we need a theoretical framework with respect to clusters in graphs. 

\subsection{\label{bad}Modularity density: The bad}
Currently, the so-called modularity density function \cite{newman2004} is most commonly used to approximate the connectedness of a subset. This function gives a value $Q$ to each partition of a network into $k$ subsets:

{\small
\begin{align}
	\label{mod}
	&Q=\sum_{i=1}^k\left[\frac{e_i}{m}-\left(\frac{d_i}{2m}\right)^2\right]
\end{align}
}

In equation (\ref{mod}) $e_i$ is the number of edges in subset $i$, $m$ corresponds with the total number of edges in the network, and $d_i$ is the total degree of nodes in subset $i$. The function holds the difference between the connectivity within the partitions or subsets on the one hand, and the expected connectivity of a random graph with the same degree sequence on the other hand \cite{molloy1995}. It is worth noting that modularity density aims to approximate a good division of a graph into communities, but that it is not explicitly defined what a community entails. Instead, the goal is to find a value such that the division into subsets is ``sufficiently'' different from the value in a similar random graph. Several downsides of the modularity density function have, however, been shown in literature \cite{good2010}:

\begin{enumerate}
\item Resolution limit: Smaller communities may be hidden within larger subsets. The modularity density function may not correctly identify small clusters because a higher value $Q$ can be obtained by reporting larger sets. In particular, the choice of a random graph as null model is problematic in this regard.
\item Degeneracy: An exponential number of high-quality solutions exist with a value $Q$ close to the optimum. These solutions may differ (greatly) from one another in terms of the reported partitions. The global optimum itself is, however, difficult to determine. Especially, in hierarchical or modular networks the multitude of possible competitive solutions increases further, without a clear manner to distinguish between them. 
\item Limiting behavior: The maximum value for $Q$ depends on the graph size $n$ and on the number of communities $k$. This results in higher modularity function values for larger networks and for networks with more modules. Hence, a high value $Q$ may indicate that the graph is very different from a random graph with a same degree sequence, rather than that it has a high modularity. 
\end{enumerate} 

In spite of these major pitfalls, modularity density is employed in most recent publications on community detection in (large) graphs. As a result, it is not apparent what the proposed techniques actually optimize. 

\subsection{Maximal cliques: The useful}
A field of research related to community detection concerns the maximal clique problem. A clique is a subgraph in which every two nodes are connected, and a maximal clique is a clique which cannot be extended by including one more adjacent vertex. The maximal clique problem focuses on detecting all maximal cliques in a graph, similar to communities. The major difference is that whereas communities have no predefined degree of connection, the maximal cliques do. Recent work on finding maximal cliques in real-life networks has illustrated that this approach leads to valuable results in terms of both memory management and algorithm development, see e.g. \cite{cheng2011}, \cite{eppstein2011} and \cite{conte2016} for some examples. Especially, the decomposition approaches of e.g. \cite{conte2016} allow for finding maximal cliques quickly in subsets of the overall network. 

Recently, \cite{pattillo2013} proposed a framework for clique relaxation models based on known clique-defining properties. Particularly interesting are the different alternative clique definitions discussed, which allow for several types of clique relaxations. They are based on the elementary properties distance, diameter, domination, degree, density and connectivity of cliques. Relaxations of each of these restrictions in cliques may prove useful for defining community structures. As an example, \cite{perkins2009} employed a clique relaxation approach to analyze the relationship between different genes in biological data.

Based on both the advances in terms of algorithms for maximal cliques and the framework for clique relaxation, we believe it would be worthwhile to employ cliques to model clusters in large graph problems. It is both interesting and somewhat strange that in literature there appears to be a dividing line between research on clique optimization and its applications in among others the field of bioinformatics (see e.g. \cite{perkins2009}) on the one hand and community detection on the other hand. This observation further highlights the need for an approach which considers both fields of research.

\subsection{Clique relaxation: The better?}
In this section, we propose an alternative to the modularity density function, derived from the clique relaxations of \cite{pattillo2013}. Before going into detail, we first discuss some useful notations and definitions. A graph $G=(V,E)$ consists of a set of nodes or vertices $V$ and a set of edges $E$, which connect pairs of nodes. Two nodes $v$ and $w$ are said to be neighbors if they share an edge, i.e. if $(v,w)\in E$. The set $N_G(v)$ contains all neighbors of $v$ in $G$, and $|N_G(v)|$ is called the degree of $v$ in $G$ denoted as $deg_G(v)$. $\delta(G)$ and $\Delta(G)$ are the minimum and maximum degree of any node in $G$ respectively. The subgraph $G[S]$, with $S\subseteq V$, is obtained by removing all nodes $V\setminus S$ and any edge connected to at least one such node from $G$. Finally, $\rho(G)$ is the density of $G$ and equals the ratio of the number of edges to the total number of possible edges: $\rho(G)=|E|/\binom {|V|}{2}$.

Since a community can be informally defined as ``a set of nodes with more connections within the set than with other nodes outside of the set'', we focus on edge connectivity for a more formal definition. Hence, we use the $(\lambda, \gamma)$-quasi-clique definition ($\lambda, \gamma \in ]0;1]$) for a set $S$, which holds if $\delta(G[S])\ge \lambda(|S|-1)$ and $\rho(G[S])\ge \gamma$. This second-order relaxed clique definition implies that, based on input values for both $\lambda$ and $\gamma$, we can impose restrictions on how connected the communities or subsets should be. Low (high) values for $\lambda$ and $\gamma$ imply a low (high) degree of connectedness. Our objective is to find all maximal $(\lambda, \gamma)$-quasi-cliques, instead of optimizing the modularity density function. 

\begin{itemize}
\item If both $\lambda$ and $\gamma$ equal one, we have the maximal clique problem, since we require that every pair of nodes in a subset have an edge between them. As a result, here a community is the same as a clique. 
\item A value for $\lambda$ between zero and one implies that each node does not have to be connected to every other node in the subset. As an example, consider a subset of a larger network with five nodes. If $\lambda$ equals 0.75 this means that the nodes in the subset only need an edge to three out of the other four vertices.
\item A value for $\gamma$ between zero and one states that the total number of edges in the subset can be smaller than the total number of possible edges. In the example, the total number of possible edges is $\frac{5!}{2!\cdot 3!}$ or 10. If the value for $\gamma$ is 0.80, a total of at least eight edges is required in a subset.
\item The combination of the values for $\lambda$ and $\gamma$ yields a combination of restrictions for the total number of edges and the number of edges per vertex. This way, the required structure of communities can be set in advance in a formal manner, and we avoid a purely black-or-white view on clusters in graphs.
\end{itemize}

Other clique relaxation definitions such as $s$-defective clique or $k$-core can be used as well, but the overall logic remains the same. A clique relaxation approach is used to decide on the structure of subsets or communities, based on one or more parameters.

Revisiting the shortcomings of the modularity density function discussed in section \ref{bad}, we can conclude the following:
\begin{enumerate}
\item Resolution limit: The quasi-clique definition states that the focus should be on finding any maximal community which satisfies the relaxed clique restrictions. Hence, there is no bias towards finding larger clusters and omitting smaller ones. Only the selected values for $\lambda$ and $\gamma$ impact the sizes of the cliques reported, and also the number of clusters identified. Finally, no comparison is made with a random graph, as is the case for modularity density.
\item Degeneracy: Since the $(\lambda, \gamma)$-quasi-clique definition optimizes the assignment of nodes to cliques, some combinations of assignment may be considered equivalent. For example, a solution with cliques with sizes 8 and 2 may be considered equivalent to a solution with cliques of size 6 and 4, since in both solutions a total of 10 nodes are assigned to cliques. Whereas it can be stated that multiple solutions exist with the same objective function value, the global optimum should be determinable in a clear manner. We conclude that the degeneracy issue warrants more investigation, even for clique relaxations.
\item Limiting behavior: The objective function value depends on the graph size $n$, since it can on average be expected that larger graphs contain more cliques. This is not a problem since, unlike for modularity density, no comparison is made with a random graph, and as a result the impact on the objective function comes purely from the network's parameters (which include but are not limited to its size $n$).
\end{enumerate}

\section{\label{issue2}What is the impact of algorithms?}
Second, it is often unclear which parts of the algorithms from literature contribute to the techniques' performance and to what extent. This way, the explanation for differences between several algorithms are not discussed, and it is difficult to determine whether one algorithm is really better than another one, let alone why. In this section, we focus on (meta)heuristic techniques since these algorithms are better suited to solve large graph problems with 10,000 or more nodes than their exact counterparts.

\subsection{Algorithm components}
To the best of our knowledge, very few if any papers on community detection (see \cite{fortunato2010} and \cite{fortunato2016} for extensive overviews) analyze the proposed techniques to show whether the composing parts are worth their salt. Nonetheless, it would prove a beneficial endeavor to discuss the individual parts of an algorithm in more detail and to demonstrate the added value of each crucial component. We do not imply that authors should investigate the impact of each part of a (well-)known metaheuristic framework for instance, but they should rather analyze the effect of any novel parts such as for instance a new local search. Statistical tests should be applied as well, to ensure that the added value can be validated. For a recent example of a research paper which tests the effects of newly introduced algorithm components, we refer to \cite{leyman2017b}, who discuss a new local search framework as part of a metaheuristic for optimizing net present value in a project scheduling context. 

One can also wonder how suitable commonly used metaheuristic frameworks are for solving large optimization problems. It would not be unreasonable to assume that if we want algorithms to scale to datasets with tens of thousands if not millions of nodes, we need to develop algorithm components with computational complexity $O(n\cdot log$ $n$) or even $O(n)$. Hence, we believe that a more thorough analysis of solution techniques and their composing parts is recommended in order to properly show their added value and (potential) shortcomings. Of interest in this regard are hyperheuristics, which could be used to construct or select heuristic techniques based on the problem instance under consideration \cite{burke2012}. This would allow the method to focus on a search space of heuristic techniques rather than solutions.

\subsection{Algorithm comparison}
Of particular concern in metaheuristic research is the performance evaluation of different techniques and the lack of a commonly used manner of doing so \cite{sorensen2015}. Especially in the field of community detection, comparisons between different approaches occur rarely. Nonetheless, it can be argued that a fair and independent comparison of approaches is needed \cite{kendall2015}. Allowable computation time is the most often employed termination criterion, but it is hardly fair to compare code of different researchers often run on different machines. Instead, the focus should be on evaluating algorithm efficiency rather than code efficiency \cite{hooker1995}, \cite{leyman2017a}. This way, misinterpretations and -representations of results can be avoided.

Another issue relates to the focus in literature on playing an up-the-wall game when proposing new techniques, i.e. the results of the new approach have to outperform the best known results on some benchmark dataset. From a scientific point of view the focus should on the contrary be on understanding why some methods perform better than others \cite{sorensen2015}. The insights gained can prove invaluable in designing new methods (as part of a hyperheuristic for instance) and in understanding the difficulty inherent to some classes of problems.

Finally, just like for evaluating the added value of algorithm components, statistical tests should be applied to show whether any reported differences are indeed valuable. 

\section{\label{issue3}And what about data?}
Third, the effects of data parameter values on algorithm performance are not analyzed. Even though both fictitious \cite{lancichinetti2009}, \cite{lancichinetti2008} and real-life \cite{leskovec2016} datasets with large variation in parameter values are used, the effect of these data parameters on algorithm performance and instance difficulty is never considered. It would, however, be worthwhile to investigate such issues in order to have an understanding of the performance of different techniques based on the input data. These insights may in turn prove valuable for guiding solution techniques in their quest for optimality, by including for instance learning mechanisms. 

We propose to use real-life data, such as those of \cite{leskovec2016}, to derive which data parameters are important in large networks, to complement existing parameters for clique optimization problems \cite{pattillo2013}. Based on these parameters, fictitious data can be generated with a larger variation in data parameter values, to allow for a broader analysis of algorithm performance. A major pitfall, however, may concern the reliability of the proposed techniques. It has been shown that there can be a possible difference between structural communities detected by algorithms and the metadata groups derived from the node characteristics \cite{hric2014}. As a result, it may be crucial to first investigate topological features of the graphs and derive a general description, before building actual algorithms.

The link with and effect of clique relaxation parameters (e.g. the $\lambda$ and $\gamma$ of section \ref{issue1}) should be investigated as well. In particular, what type of subsets are reported as cliques along with their size can allow for insights into the structure of large graphs. These insights can be used to determine important graph characteristics useful for e.g. bioinformatics.

To conclude this section, we would like to point out the need for testing algorithms on more than a handful of networks. Currently, most solution methodologies are only evaluated on a small number of instances, which implies that the results can hardly be generalized. It is this regard that the design of a sufficiently large and varied dataset is particularly crucial, to allow for a broad analyses of algorithms' performance.

\section{\label{concl}Conclusions \& future work}
We have discussed the shortcomings of community detection, and used them as stepping stones to propose a more formal framework for communities in large graphs. The more formal approach is derived from the maximal clique problem, and allows for a predefined degree of connectedness in graphs. It is argued that clique relaxations allow for the proper detection of groups of nodes in a graph, without any of the shortcomings of the commonly used modularity density function.

The issues regarding the impact of algorithms and data in the context of community detection have also been touched upon. Both issues need to be tackled in order to allow for a real step forward in research efforts on large graphs and community detection. 

In the future, we aim to further extend the proposed approach as well as test its performance with respect to the issues of community detection. We will evaluate the positive and negative aspects of the clique relaxation used, as well as consider possible extensions and different types of clique relaxations. Additionally, we will also investigate the second and third issue on community detection in large graphs in detail, by implementing the approach in different metaheuristic frameworks and analyzing the effects based on diverse datasets.

\vspace{1.0mm}
\textbf{Acknowledgments:} This research was supported by the Belgian Science Policy Office (BELSPO) in the Interuniversity Attraction Pole COMEX.



\begin{thebibliography}{1}
\bibitem{burke2012}
E.K.~Burke, M.~Hyde, G.~Kendall, G.~Ochoa, E.~\"{O}zcan, and J.R.~Woodward, A classification of hyperheuristic approaches., In M.~Gendreau and J.-Y.~Potvin (Eds.), \emph{Handbook of metaheuristics} (pp.449-468), Springer, 2012.

\bibitem{cheng2011}
J.~Cheng, Y.~Ke, A. W.-C.~Fu, J. X.~Yu and L.~Zhu, Finding maximal cliques in massive networks., \emph{ACM Transactions on Database Systems} 36(4): 1-34, 2011.

\bibitem{conte2016}
A.~Conte, R.~De Virgilio, A.~Maccioni, M.~Patrignani and R.~Torlene, Finding all maximal cliques in very large social networks., \emph{Proceedings of the 19th International Conference on Extending Database Technology}, 173-184, 2016.

\bibitem{eiben2015}
A.E.~Eiben and J.E.~Smith, \emph{Introduction to Evolutionary Computation}, Springer, 2015.

\bibitem{eppstein2011}
D.~Eppstein and D.~Strash, Listing all maximal cliques in large sparse real-world graphs., \emph{Lecture Notes in Computer Science} 6630: 364-375, 2011.

\bibitem{fortunato2010}
S.~Fortunato, Community detection in graphs., \emph{Physics Reports} 486: 75-174, 2010.

\bibitem{fortunato2016}
S.~Fortunato and D.~Hric, Community detection in networks: A user guide. \emph{Physics Reports}, 659: 1-44, 2016.

\bibitem{good2010}
B.H.~Good, Y.-A.~de Montjoye and A.~Clauset, Performance of modularity maximization in practical contexts., \emph{Physical Review E}, 81: 046106, 2010.

\bibitem{hooker1995}
J.N.~Hooker, Testing heuristics: We have it all wrong, \emph{Journal of Heuristics}, 1: 33-42, 1995.

\bibitem{hric2014}
D.~Hric, R.K.~Darst and S.~Fortunato, Community detection in networks: Structural communities versus ground truth \emph{Physical Review E}, 90: 062805, 2014.

\bibitem{kendall2015}
G.~Kendall, R.~Bai, J.~Blazewicz, P.~De Causmaecker, M.~Gendreau, R.~John, J.~Li, B.~McCollum, E.~Pesch, R.~Qu, N.~Sabar, G.~Vanden Berghe and A.~Yee, Good laboratory practice for optimization research., \emph{Journal of the Operational Research Society}, 67(4): 676-689, 2015.

\bibitem{lancichinetti2009}
A.~Lancichinetti and S.~Fortunato, Benchmarks for testing community detection algorithms on directed and weighted graphs with overlapping communities., \emph{Physical Review E}, 80: 016118, 2009.

\bibitem{lancichinetti2008}
A.~Lancichinetti, S.~Fortunato and F.~Radicchi, Benchmark graphs for testing community detection algorithms., \emph{Physical Review E}, 78: 046110, 2008.

\bibitem{leskovec2016}
J.~Leskovec and R.~Sosic, SNAP: A general-purpose network analysis and graph-mining library., \emph{ACM Transactions on Intelligent Systems and Technology}, 8(1): 1-20, 2016.

\bibitem{leyman2017a}
P.~Leyman and P.~De Causmaecker, Termination criteria for metaheuristics: Is computation time worth the time?, \emph{31st Annual Meeting of the Belgian Operational Research Society}, Brussels, Belgium, 2017.

\bibitem{leyman2017b}
P.~Leyman and M.~Vanhoucke, Capital- and resource-constrained project scheduling with net present value optimization., \emph{European Journal of Operational Research}, 256: 757-776, 2017.

\bibitem{molloy1995}
M.~Molloy and B.A.~Reed, A critical point for random graphs with a given degree sequence., \emph{Random Structured Algorithms}, 6: 161-180, 1995.

\bibitem{newman2004}
M.E.J.~Newman and M.~Girvan, Finding and evaluating community structure in networks., \emph{Physical Review E}, 69: 026113, 2004. 

\bibitem{pattillo2013}
J.~Pattillo, N.~Youssef and S.~Butenko, On clique relaxation models in network analysis., \emph{European Journal of Operational Research}, 266: 9-18, 2013. 

\bibitem{perkins2009}
A.D.~Perkins and M.A.~Langston, Threshold selection in gene co-expression networks using spectral graph theory techniques., \emph{BMC Bioinformatics}, 10 (Suppl. 11), S4, 2009. 

\bibitem{sorensen2015}
K.~S\"{o}rensen, Metaheuristics-the metaphor exposed., \emph{International Transactions in Operational Research}, 22: 3-18, 2015.

\end{thebibliography}
\end{document}